\documentclass[a4paper,showkeys,showpacs,preprint,onecolumn]{revtex4-1}
\usepackage[english]{babel}
\usepackage[utf8]{inputenc}
\usepackage{graphicx}
\usepackage{epstopdf}
\usepackage{footnote}
\usepackage{amsmath, amssymb}
\usepackage{dcolumn}
\usepackage{subfigure}
\usepackage{xcolor}

\usepackage{float}

\begin{document}

\title{\bf Parameter estimation in an anisotropic expanding spacetime}

\author{O. P. de S\'{a} Neto$^{(a)}$}
\author{I. G. da Paz$^{(b)}$}
\author{P. R. S. Carvalho$^{(b)}$}
\author{H. A. S. Costa$^{(b)}$}

\affiliation{(a) Coordena\c{c}\~{a}o de Ci\^{e}ncia da Computa\c{c}\~{a}o, Universidade Estadual do Piau\'{i}, 64202-220, Parna\'{i}ba, PI, Brazil}
\affiliation{(b) Universidade Federal do Piau\'{\i}, Departamento de F\'{\i}sica, 64049-550, Teresina, PI, Brazil}

\begin{abstract}
In this work, we investigate how the anisotropy affects the cosmological parameters estimation. Here the anisotropy is incorporated as a small gravitational disturbance. We calculate the Fisher information for both cosmological parameters  $\epsilon$ (expansion volume) and $\rho$ (expansion rate), and we show that the anisotropy introduces oscillations in the Fisher information spectrum. This implies that the estimation of the cosmological parameters is sensible to the direction of the momentum $k$ of particles. In addition, we observe that for small values of the momentum $k$ there is a substantial difference between the Fisher information spectrum for the minimum and conformal couplings.
\end{abstract}

\maketitle

\tableofcontents
 
 \section{Introduction}

Quantum metrology exploits quantum properties to improve the estimation of physical parameters. Roughly speaking, a typical quantum metrology scheme contains the following steps. First, a quantum probe state is prepared. Then, the probe state is modified by any evolution process, encoding the set of parameters to be estimated. The parameter information is then encoded in a quantum observable. Finally, the estimation of single or multi parameters is performed. The estimation protocol introduces uncertainty to the measured values \cite{Geremia}. The effect of this additional errors can be minimized by repeating the measurements and averaging the outcomes \cite{Helstrom, Holevo, Paris, Helstrom2}. In addition, it is well known that the mean variance of the uncertainty for a given measurement of a parameter $\theta$ is lower bound by the Cram\'{e}r-Rao inequality, $\mathrm{Var}(\theta) \geq 1/\sqrt{NF_{\mathrm{Q}}(\theta)}$ \cite{Braunstein, Braunstein2}, where $N$ is the number of identical measurements repeated and $F_{\mathrm{Q}}(\theta)$ is the quantum Fisher information. It is noteworthy that the quantum Fisher information is the central quantity that allow us to determine the ultimate limits of precision in the estimation of a parameter $\theta$.
 
Recently, the techniques of quantum metrology have been employed in the context of quantum field theory in curved spacetime \cite{Mehdi}. In particular, previous researches have investigated the metrology of a wide range of relativistic phenomena, including estimation of entanglement \cite{Genoni, Brida}, gravitational waves \cite{Vallisneri, Aasi,GW, Rideout}, Unruh-Hawking effect \cite{Aspachs, Hosler, Yao, Wang, Tian, Yang}, parameters of classical spacetimes \cite{Downes}, Schwarzschild spacetime parameters of the Earth \cite{Bruschi}, $\kappa$-deformation of noncommutative spacetime \cite{Liu}, and Lorentz violation parameter \cite{Costa}. In addition, parameter estimation has been studied in the context of quantum cosmology. As examples of such application of quantum metrology we can mention the estimation of the expansion parameters of the universe \cite{Wang2, Huang} and the influence of the Lorentz invariance violation on the ultimate limits of precision on the estimation of cosmological parameters \cite{Liu2}. However, these works do not take into account anisotropy effects on the cosmological parameters estimation. The estimation of cosmological parameters is one of the central issues in modern cosmology, thus, it is interesting to  investigate quantum metrology protocol in a more realistic model. 
 
 There are two central goals here: First, investigate the effects of an anisotropic perturbation of the metric on the estimation of the cosmological parameters $\epsilon$ (expansion volume) and $\rho$ (expansion rate). Second, examine how the Fisher information spectrum is modified by the type of coupling between the field and the spacetime curvature. The paper is organized as follow. In Section 2, we introduce the isotropic and anisotropic models for the expanding spacetime. In addition, we explicitly compute the Bogoliubov coefficients for both the models. In Section 3, we present a review of the basic notions of local quantum estimation theory. In Section 4, we calculate the Fisher information of the cosmological parameters, and study how the anisotropy affects the estimation scheme. Finally, Section 5 summarizes the results of this work and draws our conclusions.

 \section{Physical model}
In this section we study the isotropic and anisotropic models for the expanding spacetime. We start with the Friedmann-Robertson-Walker (FRW) model which describes a homogeneous and isotropic Universe. As it is known, under such conditions the Einstein field equations $G^{\mu}_{\nu} = \varsigma T^{\mu}_{\nu}$, where $G^{\mu}_{\nu}$ is the Einstein tensor, $\varsigma = 8\pi G/c^4$, and $T^{\mu}_{\nu}$ is the energy-momentum tensor for a perfect fluid in the comoving frame, can  be  solved  exactly,  resulting  in  the  metric 
 \begin{align*}
 ds^2 = c^2dt^2 - a^2(t)\left[\frac{dr^2}{1 - \kappa r} + r^2(d\theta^2 + \sin^2\theta d\phi^2)\right].
 \end{align*}
Here, $a(t)$ is the scalar factor and $\kappa = -1, 0, 1$ is the curvature parameter, respectively for open, flat, and closed geometry.

 \subsection{Isotropic case}
Here we study the expanding spacetime in an isotropic medium. Then, we consider a real and non-interacting scalar field $\phi$ with mass $m$ propagating in (1+1)-dimensional FRW metric with $\kappa = 0$. In terms of the conformal time $\eta = \int \frac{dt}{a(t)}$ the metric of the FRW can be written as \cite{Martinez0}
\begin{align*}
 ds^2 = a^2(\eta)\left(d\eta ^2 - dx^2\right).
\end{align*} 
From this, it is clear that the FRW spacetime is conformally flat: $g_{\mu\nu} = a^2(\eta)\eta_{\mu\nu}$.  Now, we consider the dynamics of the scalar field $\phi$ which is described by the covariant form of the Klein-Gordon equation 
 \begin{align} \label{Eq4}
 \left[\frac{1}{\sqrt{-g}}\partial_{\mu}(\sqrt{-g}g^{\mu\nu}\partial_{\nu}) + a^2(\eta)m^2\right]\phi = 0.
\end{align}
Since the spatial translation of the spacetime is invariant, we can use the method of variables to solve the Eq. (\ref{Eq4}). In the conformal observer frame it may be written  as 
 $\phi_k(\eta, x) = a^{-1}(\eta)e^{ikx}\chi_k(\eta)$, where $\chi_k(\eta)$ satisfies the following equations of motion
  \begin{align}\label{EqM}
   \left[\partial_{\eta}^2 + k^2 + a^2(\eta)m^2\right]\chi_k(\eta) = 0.
  \end{align}
Now, to obtain the solution of Eq. (\ref{EqM}) we consider the following form for the scale factor $a^2(\eta)$ \cite{Fuentes0}
\begin{equation}\label{a}
 a^2(\eta) = 1 + \epsilon(1 + \tanh(\rho\eta)),
\end{equation}
where $\epsilon$ and $\rho$ are positive real parameters controlling the total volume and the rapidity of the expansion, respectively. The scalar factor $a^2(\eta)$ defined above is a smooth function and approaches for a constant value in the distant past (the so-called in-region, where $\eta \rightarrow -\infty$) and in the far future (the out-region, where $\eta \rightarrow \infty$). Consequently, the metric above describes a flat spacetime in these asymptotic limits. As the spacetime is not stationary but possesses stationary asymptotic regions we can give a particle interpretation to the solution of the field equations in both distant past and far future. By using (\ref{a}), we can solve the Eq. (\ref{EqM}) in terms of hypergeometric functions. The two set of solutions in the in- and out- regions are given by
\begin{align}
 \chi_k^{\mathrm{in}}(\eta) &= \frac{e^{-i\omega_{+}\eta - \frac{i\omega_{-}}{\rho}\ln[2\cosh(\rho\eta)]}}{\sqrt{4\pi\omega_{\mathrm{in}}}} F_1\Bigg[1 + \frac{i\omega_{-}}{\rho}, \frac{i\omega_{-}}{\rho}, 1 - \frac{i\omega_{\mathrm{in}}}{\rho}, \frac{1}{2}(1 + \tanh(\rho\eta))\Bigg], \nonumber  \\
 \chi_k^{\mathrm{out}}(\eta) &= \frac{e^{-i\omega_{+}\eta - \frac{i\omega_{-}}{\rho}\ln[2\cosh(\rho\eta)]}}{\sqrt{4\pi\omega_{\mathrm{out}}}} F_1\Bigg[1 + \frac{i\omega_{-}}{\rho}, \frac{i\omega_{-}}{\rho}, 1 + \frac{i\omega_{\mathrm{out}}}{\rho}, \frac{1}{2}(1 - \tanh(\rho\eta))\Bigg], \label{chi}
\end{align}
 where the $F_1$ are hypergeometric functions, and for notational convenience we have defined
\begin{align*}
 \omega_{\mathrm{in}} &= \sqrt{k^2 + m^2}, \quad \omega_{\mathrm{out}} = \sqrt{k^2 + (1 + 2\epsilon)m^2}, \\ \omega_{+} &= \frac{1}{2}(\omega_{\mathrm{out}} + \omega_{\mathrm{in}}), \quad \omega_{-} = \frac{1}{2}(\omega_{\mathrm{out}} - \omega_{\mathrm{in}}).
\end{align*}
Notice that in the limit $\eta \rightarrow -\infty$, the first solution in (\ref{chi}) can be approximated for $$\chi_k^{\mathrm{in}}(\eta \rightarrow -\infty) \rightarrow \frac{1}{\sqrt{4\pi\omega_{\mathrm{in}}}}e^{-i\omega_{\mathrm{in}}\eta},$$ and by taking the limit $\eta \rightarrow \infty$, the second solution in (\ref{chi}) becomes $$\chi_k^{\mathrm{out}}(\eta \rightarrow \infty) \rightarrow \frac{1}{\sqrt{4\pi\omega_{\mathrm{out}}}}e^{-i\omega_{\mathrm{out}}\eta}.$$     

The asymptotic solutions above are connected via a Bogoliubov transformation. This can be done by using some basic properties of hypergeometric functions. For example, any hypergeometric function can be written as the linear combination 

\begin{align*}
F_1[a,b;c;z] &= \frac{\Gamma(c)\Gamma(c-a-b)}{\Gamma(c-a)\Gamma(c-b)}F_1[a,b;a+b+1-c; 1-z] \nonumber \\
& + \frac{\Gamma(c)\Gamma(a+b-c)}{\Gamma(a)\Gamma(b)}(1-z)^{c-a-b}F_1[c-a,c-b;1+c-a-b; 1-z]. 
\end{align*}
Another property of hypergeometric functions is the relation
\begin{align*}
F_1[a,b;c;z] = (1 - z)^{c-a-b}F_1[c-a,c-b;c;z].
\end{align*}
By using the above properties, we can write $\chi_k^{\mathrm{in}}(\eta)$ as   
\begin{align}
 \chi_k^{\mathrm{in}}(\eta) = \alpha_{k}^{\text{iso}} \chi_k^{\mathrm{out}}(\eta) + \beta_{k}^{\text{iso}} \chi_{k}^{\mathrm{out}*}(\eta),
\end{align}
where 

\begin{align}
  \alpha_{k}^{\text{iso}} &= \sqrt{\frac{\omega_{\mathrm{out}}}{\omega_{\mathrm{in}}}}\frac{\Gamma\Big(1 - \frac{i\omega_{\mathrm{out}}}{\rho}\Big)\Gamma\Big(-\frac{i\omega_{\mathrm{in}}}{\rho}\Big)}{\Gamma\Big(-\frac{i\omega_{+}}{\rho}\Big)\Gamma\Big(1 - \frac{i\omega_{+}}{\rho}\Big)}, \\
  \beta_{k}^{\text{iso}} &= \sqrt{\frac{\omega_{\mathrm{out}}}{\omega_{\mathrm{in}}}}\frac{\Gamma\Big(1 - \frac{i\omega_{\mathrm{out}}}{\rho}\Big)\Gamma\Big(\frac{i\omega_{\mathrm{in}}}{\rho}\Big)}{\Gamma\Big(\frac{i\omega_{-}}{\rho}\Big)\Gamma\Big(1 + \frac{i\omega_{-}}{\rho}\Big)}. \label{ab1}
\end{align}
Here $\alpha_{k}^{\text{iso}}$ and $\beta_{k}^{\text{iso}}$ are the Bogoliubov coefficients for transition from the out-state to the in-state.
 

\subsection{Anisotropic case}
Here we study the dynamics of a massive scalar field $\phi(x)$ in an expanding spacetime with anisotropy \cite{Pierini}. Let's start with the Lagrangian density for the scalar field in a general curved spacetime  
\begin{align} \label{L}
 \mathcal{L} &= \frac{1}{2}\sqrt{-g}[g^{\mu\nu}\partial_{\mu}\phi\partial_{\nu}\phi - (m^2 + \xi R)\phi^2],
\end{align}
where $g^{\mu\nu}$ is the metric with determiant $g$, $R$ is the Ricci scalar curvature, and $\xi$ is a dimensionless parameter which characterizes the coupling. We emphasize two interesting values for the coupling $\xi$, i.e.,  $\xi = 0$ and $\xi = 1/6$. On the one hand, for $\xi = 0$, the field is said to be minimally coupled to the metric. On the other hand, for $\xi = \frac{1}{6}$, (\ref{L}) is conformally invariant in the massless limit. Again, the dynamics of the field operator $\phi$ is governed by the covariant Klein-Gordon equation (\ref{Eq4}). In particular, let us consider a specific model of an anisotropic universe, a Bianchi Type I spacetime, with line element
\begin{align}\label{metric1}
ds^2 = dt^2 - \sum_{j=1}^3 a^2_j(t)dx^2_j,
\end{align} 
where $a_j(t)$ are arbitrary functions of time. Despite it describes a spatially homogeneous universe it is a non-isotropic toy model of the universe. As was done in Refs. \cite{Birrell01, Birrell02, Zeldovich}, we consider the scale factors as $a_j(t) = 1 + h_j(t)$ where $h_j(t)$ is assumed to be small and it is treated as a parturbation, i.e., $\mathrm{max}|h_j(t)| \ll 1$. In terms of the conformal time coordinate $d\eta = a^{-1}(t)dt$, the metric (\ref{metric1}) reads
  \begin{align}\label{metric2}
ds^2 = a^2(\eta)\Bigg[d\eta^2 - \sum_{j=1}^3(1 + h_j(\eta))dx_j^2\Bigg]. 
\end{align} 
  As an example, let us now assume $h_j(\eta)$ to be 
  \begin{align}
  h_j(\eta) = e^{-\rho\eta^2}\cos(\epsilon\eta^2 + \delta_j),
  \end{align}
  with $\delta_j = \frac{\pi}{2}, \frac{\pi}{2} + \frac{2\pi}{3}, \frac{\pi}{2} + \frac{4\pi}{3}$. It is worth to mention that this choice satisfies the condition $\sum_{j=1}^3h_j(\eta) = 0$. Then, the dynamics of the scalar field $\phi$ in the conformal observer frame  is given by 
 \begin{align} \label{EqAniso}
 \partial_{\eta}^2\phi + 2\frac{a'(\eta)}{a(\eta)}\partial_{\eta}\phi - \nabla^2\phi + a^2(\eta)(m^2 + \xi R)\phi = 0,
\end{align}
 where the prime denotes derivatives with respect to conformal time. Because of spatial translation invariance of the spacetime, the solution of the Eq. (\ref{EqAniso}) may be written as 
\begin{align}
  \phi_k(x) = (2\pi)^{-\frac{3}{2}}a^{-1}(\eta)e^{i\bold{k}\cdot\bold{x}}f_{k}(\eta),
\end{align}
where leading order in $h_j$, $f_{k}(\eta)$ satisfies
\begin{align}
\left[\eta^{\mu\nu}\partial_{\mu}\partial_{\nu} + m^2\right]f_{k}(\eta) + V(\eta)f_{k}(\eta) = 0, \label{Eqf}
\end{align}
with 
\begin{align}
V(\eta) &= [a^2(\eta) - a^2(-\infty)]m^2 + (\xi - \frac{1}{6})a^2(\eta)R(\eta) - \sum_{j=1}^3h_j(\eta)k_j^2 . \nonumber
\end{align}
Now, by imposing the following conditions \cite{Birrell01, Birrell02}
\begin{align} \label{Conditions}
\begin{split}
 a^2(\eta) R(\eta) &\to 0  \,\, \text{as} \,\, \eta \to \pm \infty \,\, \text{if} \,\, \xi \ne 1/6,  \\
 h_j(\eta) &\to 0 \,\, \text{as} \,\, \eta \to \pm \infty, \\
 a^2(\eta) &\to a^2(\pm \infty) < \infty \,\, \text{as} \,\, \eta \to \pm \infty \,\, \text{if} \,\, m \ne 0,
 \end{split}
\end{align}
one may treat $V(\eta)$ as small and solve the Eq. (\ref{Eqf}) by iteration to the lowest order in $V(\eta)$ in terms of the momentum space propagator.  In this case, the integral form of the Eq. (\ref{Eqf}) becomes
\begin{align}\label{fk}
 f_{k}(\eta) = f^{\mathrm{in}}_{k}(\eta) - \int^{\infty}_{-\infty} G_{r}(\eta,\eta')V(\eta')f_{k}(\eta')d\eta' , 
\end{align}
where $f^{\mathrm{in}}_{k}(\eta)$ is the free-wave solution propagating from the in-region, which is defined by $f^{\mathrm{in}}_{k}(\eta) = (2\omega)^{-\frac{1}{2}}e^{-i\omega\eta}$ with $\omega = \sqrt{k^2 + m^2}$. The propagator $G_{r}(\eta,\eta')$ satisfies
\begin{align}
\left[\eta^{\mu\nu}\partial_{\mu}\partial_{\nu} + m^2\right]G_{r}(\eta,\eta') = \delta(\eta - \eta')\, ,
\end{align}
and, in momentum space, it reads
\begin{align}
   G_{r}(\eta,\eta') = \frac{1}{2\pi}\int\frac{e^{-ik'_{0}(\eta' - \eta)}}{k'^{2}_{0} - \omega^{2}_{k} - i\varepsilon}dk'_{0}\, . 
\end{align}
 The momentum integral can be performed by closing the integration contour in the upper-half complex momentum plane. However, for the calculation of the Bogoliubov coefficients, it suffices to notice that in the limit $\eta \longrightarrow \infty$, $f_{k}(\eta)$ can be written in terms of the mode functions $f^{\mathrm{in}}_{k}(\eta)$ as
\begin{align}
f^{\mathrm{out}}_{k}(\eta) = (2\omega)^{-\frac{1}{2}}\left[\alpha_{k}^{\text{aniso}} e^{-i\omega\eta} + \beta_{k}^{\text{aniso}} e^{i\omega\eta}\right], \label{fklater}
\end{align}
 where the resulting Bogoliubov coefficients are given
\begin{align}
\alpha_{k}^{\text{aniso}} &= 1 + \frac{i}{\sqrt{2\omega}}\int^{\infty}_{-\infty}e^{i\omega\eta'}V(\eta')f_{k}(\eta')d\eta', \nonumber \\
\beta_{k}^{\text{aniso}} &= -\frac{i}{\sqrt{2\omega}}\int^{\infty}_{-\infty}e^{-i\omega\eta'}V(\eta')f_{k}(\eta')d\eta'. \label{Alpbeta}
\end{align}
 To the lowest order in $V(\eta)$ we have $f_{k}(\eta) \cong f^{\mathrm{in}}_{k}(\eta)$, and theBogoliubov coefficients become
\begin{align}
\alpha_{k}^{\text{aniso}} &= 1 + \frac{i}{2\omega}\int^{\infty}_{-\infty}V(\eta')d\eta', \nonumber \\
\beta_{k}^{\text{aniso}} &= -\frac{i}{2\omega}\int^{\infty}_{-\infty}e^{-2i\omega\eta'}V(\eta')d\eta'. \label{Alpbeta2}
\end{align}
Now, let us consider a scale factor (\ref{a}) which allows us to calculate the Bogoliubov coefficients and satisfy the conditions (\ref{Conditions}). Note that $a^2(\eta)$ represents conveniently a spacetime that undergoes a period of smooth expansion and becomes flat in the distant past and in the far future. By inserting in (\ref{Alpbeta2}) the explicit form of $V(\eta)$ and $a^2(\eta)$, we can write $\alpha_k^{\text{aniso}}$ and $\beta_{k}^{\text{aniso}}$ as
\begin{align} \label{betak}
\begin{split}
\alpha_k^{\text{aniso}} &= 1 + \alpha_k^{(m)} + \alpha_k^{(\xi)} + \alpha_k^{(h)}, \\
\beta_k^{\text{aniso}} &= \beta_{k}^{(m)} + \beta_{k}^{(\xi)} + \beta_{k}^{(h)},
\end{split}
\end{align}
where to leading order in $\epsilon$ we find 
\begin{align}
\alpha_k^{(m)} &= \frac{i\epsilon m^2}{\omega}, \\
\alpha_k^{(\xi)} &= (\xi - \frac{1}{6})\frac{i\rho}{12\epsilon\omega}[(\epsilon + 1)\ln(2\epsilon + 1) - 2\epsilon], \\
\alpha_k^{(h)} &= \frac{i\sqrt{\pi}}{2\omega}\sum_{j=1}^3k_j^2\mathrm{Re}\Bigg(\frac{e^{-i\delta_j }}{\sqrt{\rho + i\epsilon})}\Bigg), \\
\beta_{k}^{(m)} &= -\frac{m^2\epsilon}{2\omega\rho}\frac{\pi}{\sinh(\frac{\pi\omega}{\rho})}, \\
\beta_{k}^{(\xi)} &= (\xi - \frac{1}{6})\frac{\omega\epsilon}{6\rho}\frac{\pi}{\sinh(\frac{\pi\omega}{\rho})}, \\
\beta_{k}^{(h)} &= -\frac{\sqrt{\pi}}{2\omega}\sum_{j=1}^3k_j^2\mathrm{Re}\Bigg(\frac{e^{-i\delta_j + i\frac{\pi}{2} -\frac{\omega^2}{\rho + i\epsilon}}}{\sqrt{\rho + i\epsilon})}\Bigg).
\end{align}


\section{Methods}
In this section, let us calculate the quantum Fisher information matrix (QFIM) \cite{Liu3, Safranek}. For this, suppose a vector of parameters $\overrightarrow{x} = (x_1, x_2, ..., x_N,...)^T$ where $x_N$ is the Nth parameter. The vector of parameters is encoded in the density matrix $\hat{\rho} = \hat{\rho}(\overrightarrow{x})$. Thus, the QFIM is defined as
  \begin{align} \label{QFIM}
  \mathcal{H}_{ab} = \frac{1}{2}\mathrm{Tr}[\hat{\rho}\{\hat{L}_a, \hat{L}_b\}],
\end{align}
  where $\hat{L}_a$ and $\hat{L}_b$ are the symmetric logarithmic derivative (SLD) for the parameters $x_a$ and $x_b$, respectively. $\hat{L}_a$ and $\hat{L}_b$ are obtained via the equation
  \begin{align}
   \partial_{a(b)}\rho = \frac{1}{2}[\hat{L}_{a(b)}\hat{\rho} + \hat{\rho}\hat{L}_{a(b)}].
   \end{align}
  Note that the SLD operators are Hermitian operators with expected value $\mathrm{Tr}(\hat{L}_{a(b)}) = 0$. By using the equation above, $\mathcal{H}_{ab}$ can also be expressed by 
 \begin{align}
  \mathcal{H}_{ab} = \mathrm{Tr}[\hat{L}_b\partial_a\hat{\rho}] = -\mathrm{Tr}[\hat{\rho}\partial_a\hat{L}_b].
 \end{align}
 From Eq. (\ref{QFIM}), the diagonal entries of QFIM are $\mathcal{H}_{aa} = \mathrm{Tr}[\hat{\rho}\hat{L}_a^2]$ and $\mathcal{V}_{bb} = \mathrm{Tr}[\hat{\rho}\hat{L}_b^2]$. In particular, if we consider the density matrix with the spectral decomposition $\hat{\rho} = \sum_{i=0}^{d-1}\lambda_i |\lambda_i\rangle\langle\lambda_i|$, the QFIM can be evaluated as
 \begin{align}
  \mathcal{H}_{ab} = \sum_{i,j=0,\lambda_i+\lambda_j \neq 0}^{d - 1}\frac{2\mathrm{Re}[\langle\lambda_i|\partial_a\hat{\rho}|\lambda_j\rangle\langle\lambda_j|\partial_b\hat{\rho}|\lambda_i\rangle]}{\lambda_i + \lambda_j},
 \end{align}
 where $\lambda_i$ and $|\lambda_i\rangle$ are the eigenvalues and the corresponding eigenstates, respectively. It is usually assume that $\lambda_i > 0$ and $0 \leq i \leq d - 1$ where $d := \mathrm{dim}\rho$ is the dimension of the density matrix. By substituting the spectral decomposition of $\hat{\rho}$ into the equation above, it can be rewritten as
 \begin{align}
 \mathcal{H}_{ab} = \sum_{i=0}^{d - 1}\frac{(\partial_a\lambda_i)(\partial_b\lambda_i)}{\lambda_i} + \sum_{i\neg j,\lambda_i+\lambda_j \neq 0} \frac{2(\lambda_i - \lambda_j)^2}{\lambda_i + \lambda_j}\mathrm{Re}[\langle\lambda_i|\partial_a\lambda_j\rangle\langle\partial_b\lambda_j|\lambda_i\rangle]
 \end{align}
 The first term in the equation above can be viewed as the counterpart of the classical Fisher information as it only contains the derivatives of the eigenvalues and can be regarded as the counterpart of the probability distribution. On the other hand, the second term is purely quantum.

 \section{Results}
 
\subsection{Isotropic case}
  
 Let us consider that the field is in the vacuum state in the asymptotic past $|0\rangle_{\mathrm{in}}$, which is annihilated by $\hat{a}_{\textbf{k}}^{\mathrm{in}}$. We shall restrict our analysis to the $k$, $-k$ sector of the density matrix for a fixed wavenumber $k$, since the Lagrangian density we consider here will only mix modes of opposite momenta. This mixing of positive- and negative-frequency modes of the same wavenumber by the dynamics of spacetime leads to the excitation of the vacuum fluctuations, and consequently to particle creation. Thus, we would like to know the form of the state $|0\rangle_{\mathrm{in}}$ in the asymptotic future. To achieve this goal, let's begin with the fact that 
\begin{align*}
\hat{a}_{\textbf{k}}^{\mathrm{in}}|0\rangle_{\mathrm{in}} = 0.
\end{align*} 
 By using the Bogoliubov transformation $\hat{a}_{\textbf{k}}^{\mathrm{in}} = \alpha_k^{\text{iso}} \hat{a}_{\textbf{k}}^{\mathrm{out}} + \beta_k^{\text{iso}} \hat{a}_{-\textbf{k}}^{\mathrm{out}\dagger}$, we obtain
  \begin{align} \label{Eqinout}
  (\alpha_k^{\text{iso}} \hat{a}_{\textbf{k}}^{\mathrm{out}} + \beta_k^{\text{iso}} \hat{a}_{-\textbf{k}}^{\mathrm{out}\dagger})|0\rangle_{\mathrm{in}} = 0.
  \end{align}
  Without loss of generality, let us consider the following ``ansatz" for the vacuum state in the out-region
  \begin{align}
  |0\rangle_{\mathrm{in}} = \sum_{n = 0}^{\infty}A_n|n_k, n_{-k}\rangle_{\mathrm{out}}.
\end{align}
   By substituting the expression above into Eq. (\ref{Eqinout}), we get the recurrence relation $A_n = \left(\frac{\beta_k^{\text{iso}*}}{\alpha_k^{\text{iso}*}}\right)A_0$ and from the normalization condition we find $|A_0|^2 = 1 - \left|\frac{\beta_k^{\text{iso}}}{\alpha_k^{\text{iso}}}\right|^2$. Therefore, the vacuum state $|0\rangle_{\mathrm{in}}$ in terms of ``out" modes in the far future becomes
    \begin{align}
    |0\rangle_{\mathrm{in}} = \sqrt{1 - \gamma_{k}}\sum_{n = 0}^{\infty}\gamma_{k}^n|n_{k}, n_{-k}\rangle , \label{0k}
    \end{align}
    where
    \begin{align} \label{Gammak}
    \gamma_{k} = \left|\frac{\beta_k^{\text{iso}}}{\alpha_k^{\text{iso}}}\right|^2 = \frac{\sinh^2(\frac{\pi}{\rho}\omega_{-})}{\sinh^2(\frac{\pi}{\rho}\omega_{+})}.
    \end{align}
    Since we are working with a single mode, we will drop the frequency index $k$. Notice that the state in Eq. (\ref{0k}) is a two mode squeezed state. Let us suppose that an inertial observer in the out-region has no access to modes $-k$, thus such observer must trace over the inaccessible modes. This means that there is an unavoidable loss of information about the state of particles with modes $-k$, which essentially results in the detection of a mixed state. Therefore, an inertial observer in the out-region detects a distribution of particles with modes $k$ according to the marginal state
    \begin{align}
 \hat{\rho}_{k} &= \mathrm{Tr}_{-k}[\hat{\rho}_{k, -k}], \nonumber \\
 & = (1 - \gamma_{k})\sum_{n = 0}^{\infty}\gamma_{k}^n|n_k\rangle\langle n_k|,
\end{align}
   where the density matrix of the whole state corresponds to $\hat{\rho}_{k,-k} = |0\rangle_{\mathrm{in}}\langle 0|$. In the next section, we would use the state $\hat{\rho}_k$ as a probe state, which represents a class of signals. We are interested in estimating the cosmological parameters ($\epsilon, \rho$) with the minimum variance, which is the ultimate bound of precision imposed by the quantum theory.

  In order to estimate the cosmological parameters $\epsilon$ and $\rho$, we have evaluated the QFIM of the particle state $\hat{\rho}_k$, i.e., we calculate
  \begin{align} \label{F0}
  \mathcal{H}_{ab}^{\text{iso}} = \frac{1}{2}\mathrm{Tr}[\hat{\rho}_k L_{a}L_{b}], \quad \text{with} \quad a, b = \epsilon, \rho.
  \end{align}
  Since the density matrix $\hat{\rho}_k$ has a spectral decomposition, Eq. (\ref{F0}) can be rewritten as
  \begin{align} \label{F}
  \mathcal{H}_{ab}^{\text{iso}} = \sum_{n = 0}^{\infty}\frac{\left(\partial_a\lambda_n\right)\left(\partial_b\lambda_n\right)}{\lambda_n} + 2\sum_{n\neq m = 0}^{\infty}\frac{(\lambda_m - \lambda_n)^2}{\lambda_m + \lambda_n} \mathrm{Re}[\langle n_k|\partial_a m_k\rangle\langle \partial_b m_k|n_k\rangle],
  \end{align}
  where $\lambda_n = (1 - \gamma_k)\gamma_k^n$. Note that the last term that contains the truly quantum contribution does not contribute to $\mathcal{H}_{ab}^{\text{iso}}$ because $|n_k\rangle$ and $|m_k\rangle$ with $n\neq m$ locate on different subspaces, i.e., $\langle n_k|\partial_a m_k\rangle = \langle \partial_b m_k|n_k\rangle = 0$ $\forall n, m$.  In particular, the classical part of the quantum Fisher information $\mathcal{H}_{ab}^{\text{iso}}$ can be calculated by introducing the Eq. (\ref{Gammak}). The analytical expression of $\mathcal{F}_{\epsilon\epsilon}^{\text{iso}}$ and $\mathcal{F}_{\rho\rho}^{\text{iso}}$ can be obtained as follow. The $\mathcal{F}_{\epsilon\epsilon}^{\text{iso}}$ is given by
  \begin{align}
  \mathcal{F}_{\epsilon\epsilon}^{\text{iso}} &= \frac{1}{1 - \gamma_k}\sum_{n=0}^{\infty}\frac{1}{\gamma_k^n}\left[\partial_{\epsilon}(1 - \gamma_k)\gamma_k^n\right]^2, \nonumber \\
  &= \frac{(\partial_{\epsilon}\gamma_k)^2}{1 - \gamma_k}\sum_{n=0}^{\infty}\gamma_k^n - 2\partial_{\epsilon}\gamma_k\sum_{n=0}^{\infty}\partial_{\epsilon}\gamma_k^n + (1 - \gamma_k)\sum_{n=0}^{\infty}\frac{(\partial_{\epsilon}\gamma_k^n)^2}{\gamma_k^n}.
  \end{align}
Using the following relations
   \begin{align} 
   \partial_{\epsilon}\gamma_k &= \frac{\pi m^2}{\omega^{\mathrm{out}}_k\rho}\left[\frac{1}{\tanh(\frac{\pi\omega_{-}}{\rho})} - \frac{1}{\tanh(\frac{\pi\omega_{+}}{\rho})}\right]\gamma_k, \nonumber \\
   \partial_{\epsilon}\gamma_k^n &= \frac{\pi m^2}{\omega^{\mathrm{out}}_k\rho}\left[\frac{1}{\tanh(\frac{\pi\omega_{-}}{\rho})} - \frac{1}{\tanh(\frac{\pi\omega_{+}}{\rho})}\right]n\gamma_k^n,  \label{D1}
   \end{align}
 one can easily obtain the following result
  \begin{align} \label{F2}
  \mathcal{F}_{\epsilon\epsilon}^{\text{iso}}  = \frac{\gamma_k}{(1 - \gamma_k)^2}\frac{\pi^2 m^4}{\omega^{\mathrm{out}2}_k\rho^2}\left[\frac{\tanh(\frac{\pi\omega_{+}}{\rho}) - \tanh(\frac{\pi\omega_{-}}{\rho})}{\tanh(\frac{\pi\omega_{-}}{\rho})\tanh(\frac{\pi\omega_{+}}{\rho})}\right]^2.
  \end{align}
  Similarly, the Fisher information $\mathcal{F}_{\rho\rho}^{\text{iso}}$ is given by
  \begin{align}
  \mathcal{F}_{\rho\rho}^{\text{iso}} &= \frac{1}{1 - \gamma_k}\sum_{n=0}^{\infty}\frac{1}{\gamma_k^n}\left[\partial_{\rho}(1 - \gamma_k)\gamma_k^n\right]^2 \nonumber \\
  &= \frac{(\partial_{\rho}\gamma_k)^2}{1 - \gamma_k}\sum_{n=0}^{\infty}\gamma_k^n - 2\partial_{\rho}\gamma_k\sum_{n=0}^{\infty}\partial_{\rho}\gamma_k^n + (1 - \gamma_k)\sum_{n=0}^{\infty}\frac{(\partial_{\rho}\gamma_k^n)^2}{\gamma_k^n}.
  \end{align}
 Now, by using the relations
   \begin{align}
   \partial_{\rho}\gamma_k &= -\frac{\pi}{\rho^2}\left[\frac{\omega_{-}}{\tanh(\frac{\pi\omega_{-}}{\rho})} - \frac{\omega_{+}}{\tanh(\frac{\pi\omega_{+}}{\rho})}\right]\gamma_k, \nonumber \\
   \partial_{\rho}\gamma_k^n &= -\frac{\pi}{\rho^2}\left[\frac{\omega_{-}}{\tanh(\frac{\pi\omega_{-}}{\rho})} - \frac{\omega_{+}}{\tanh(\frac{\pi\omega_{+}}{\rho})}\right]n\gamma_k^n, \label{D2}
   \end{align} 
  we find
  \begin{align} \label{F3}
  \mathcal{F}_{\rho\rho}^{\text{iso}} = \frac{\gamma_k}{(1 - \gamma_k)^2}\frac{\pi^2}{\rho^4}\left[\frac{\omega_{-}\tanh(\frac{\pi\omega_{+}}{\rho}) - \omega_{+}\tanh(\frac{\pi\omega_{+}}{\rho})}{\tanh(\frac{\pi\omega_{-}}{\rho})\tanh(\frac{\pi\omega_{+}}{\rho})}\right]^2.
  \end{align}

\begin{figure}[H]
\centering
\includegraphics[scale=0.5]{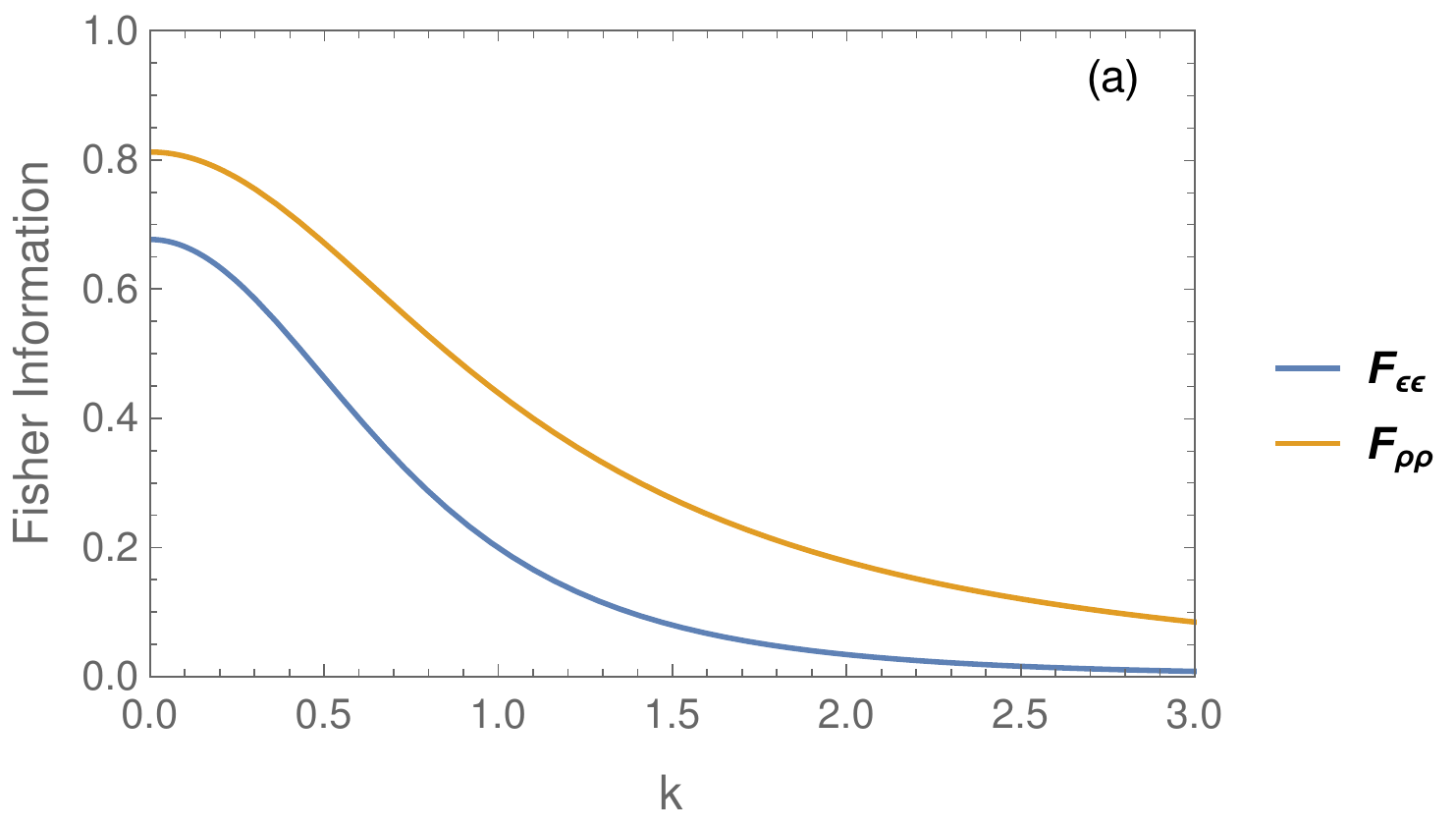} \includegraphics[scale=0.5]{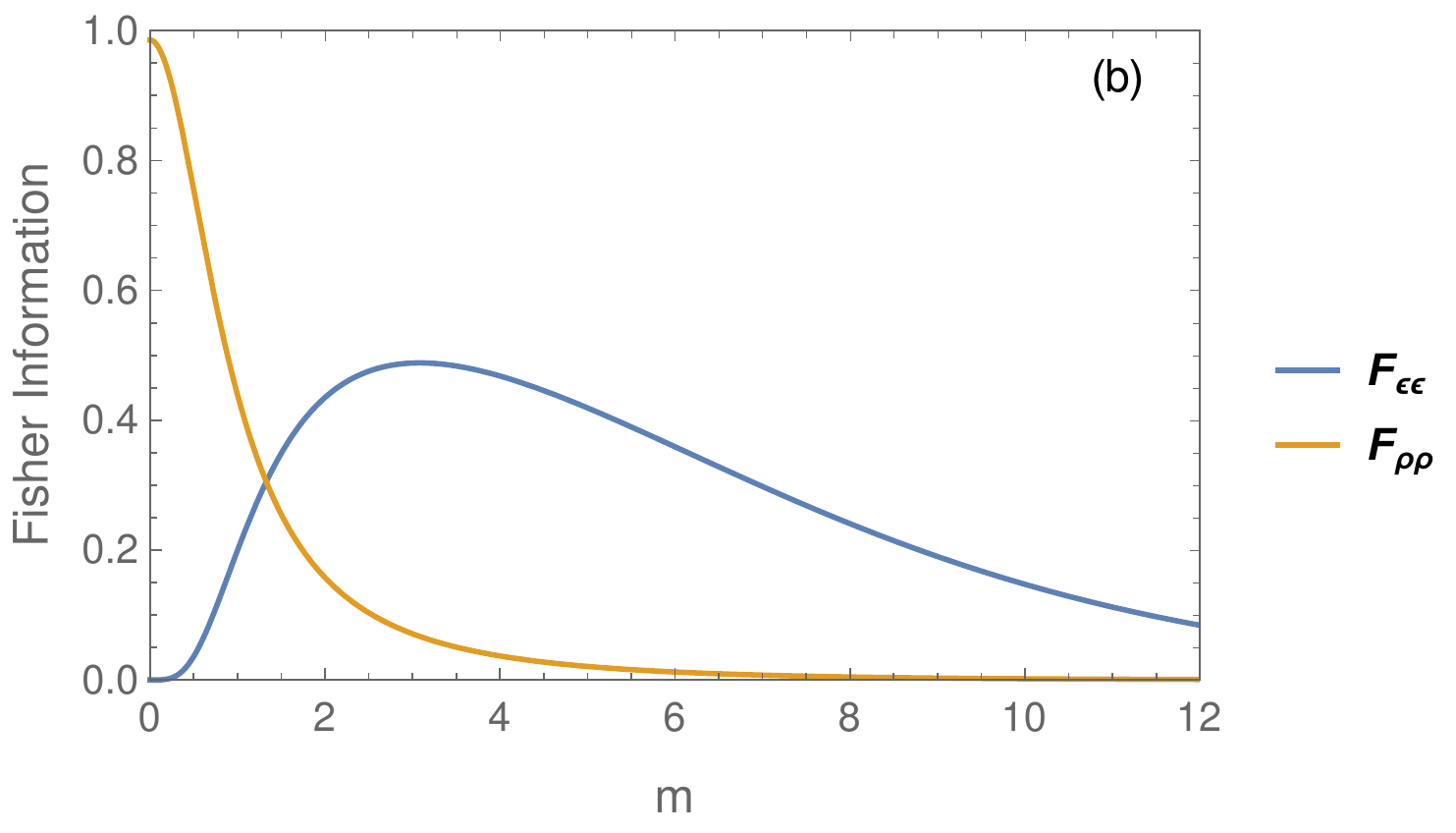}
\caption{ (a) Fisher information $\mathcal{F}_{\epsilon\epsilon}^{\text{iso}}$ and $\mathcal{F}_{\rho\rho}^{\text{iso}}$ as a function of momentum $k$, with $\rho = 15$, $\epsilon = 0.1$ and $m = 1$. (b) Fisher information $\mathcal{F}_{\epsilon\epsilon}^{\text{iso}}$ and $\mathcal{F}_{\rho\rho}^{\text{iso}}$ as a function of mass $m$, with $\rho = 15$, $\epsilon = 0.1$ and $k = 1$.  }\label{Fiso}
\end{figure}

 Note that the Fisher information $\mathcal{F}_{\epsilon\epsilon}^{\text{iso}}$ and $\mathcal{F}_{\rho\rho}^{\text{iso}}$ quantify the error in the estimation of the individual cosmological parameters. Fig. \ref{Fiso} (a) shows the Fisher information $\mathcal{F}_{\epsilon\epsilon}^{\text{iso}}$ and $\mathcal{F}_{\rho\rho}^{\text{iso}}$ as a function of the momentum of the mode $k$. We can see that $\mathcal{F}_{\epsilon\epsilon}^{\text{iso}}$ and $\mathcal{F}_{\rho\rho}^{\text{iso}}$ are  monotonic decreasing function of the momentum of the mode $k$. In this case, notice that the spectrum of the Fisher information $\mathcal{F}_{\epsilon\epsilon}^{\text{iso}}$ and $\mathcal{F}_{\rho\rho}^{\text{iso}}$ are similar. Moreover, notice that the highest precision in the estimation of both the parametrs $\epsilon$ and $\rho$ can be achieved for small momentum $k$.

 In Fig. \ref{Fiso} (b) we plot the Fisher information $\mathcal{F}_{\epsilon\epsilon}^{\text{iso}}$ and $\mathcal{F}_{\rho\rho}^{\text{iso}}$ with respect to the mass $m$. Fig. \ref{Fiso} (b) shows that the Fisher information $\mathcal{F}_{\epsilon\epsilon}^{\text{iso}}$ reaches the maximum at the optimal value of mass $m \approx 3.0$. This implies that the optimal precision in the estimation of the expansion volume  can be achieved at certain mass of particles in the measuring process. On the other hand, notice that $\mathcal{F}_{\rho\rho}^{\text{iso}}$ is a monotonic decreasing function of mass, which means that when $m \rightarrow 0$, the estimation of the expansion rate reaches the highest precision.

\subsection{Anisotropic case}

 In this section, we investigate the influence of the anisotropic perturbation on the estimation of the cosmological parameters $\epsilon$ and $\rho$. In addition, we consider the Fisher information for two values of $\xi$, namely, the weak coupling ($\xi = 0$) and the conformal coupling ($\xi = 1/6$). For this purpose, let us calculate the general analytical expression for the entries of $\mathcal{F}_{ab}^{\text{aniso}}$ associated with $\hat{\rho}_k$:
 
  \begin{align*}
  \mathcal{F}_{ab}^{\text{aniso}} &= \frac{1}{1 - \Lambda_k}\sum_{n=0}^{\infty}\frac{1}{\Lambda_k^n}\left[\partial_{a}(1 - \Lambda_k)\Lambda_k^n\right]\left[\partial_{b}(1 - \Lambda_k)\Lambda_k^n\right] \nonumber \\
  &= \frac{(\partial_{a}\Lambda_k)(\partial_{b}\Lambda_k)}{1 - \Lambda_k}\sum_{n=0}^{\infty}\Lambda_k^n - \partial_{a}\Lambda_k\sum_{n=0}^{\infty}\partial_{b}\Lambda_k^n - \partial_{b}\Lambda_k\sum_{n=0}^{\infty}\partial_{a}\Lambda_k^n + (1 - \Lambda_k)\sum_{n=0}^{\infty}\frac{(\partial_{a}\Lambda_k^n)(\partial_{b}\Lambda_k^n)}{\Lambda_k^n},
  \end{align*}
 where $\Lambda_k = \left|\frac{\beta_k^{\text{aniso}}}{\alpha_k^{\text{aniso}}}\right|^2$. By replacing the following relations in the expression above
   \begin{align} 
   \partial_{a(b)}\Lambda_k &= \Lambda_k \left[\frac{\partial_{a(b)}\beta_k^{\text{aniso}}}{\beta_k^{\text{aniso}}} + \frac{\partial_{a(b)}\beta_k^{\text{aniso}*}}{\beta_k^{\text{aniso}*}} - \frac{\partial_{a(b)}\alpha_k^{\text{aniso}}}{\alpha_k^{\text{aniso}}} - \frac{\partial_{a(b)}\alpha_k^{\text{aniso}*}}{\alpha_k^{\text{aniso}*}}\right], \nonumber \\
   &= \Lambda_k \partial_{a(b)}[\ln \Lambda_k], \nonumber \\
   \partial_{a(b)}\Lambda_k^n &= n\Lambda_k^n \left[\frac{\partial_{a(b)}\beta_k^{\text{aniso}}}{\beta_k^{\text{aniso}}} + \frac{\partial_{a(b)}\beta_k^{\text{aniso}*}}{\beta_k^{\text{aniso}*}} - \frac{\partial_{a(b)}\alpha_k^{\text{aniso}}}{\alpha_k^{\text{aniso}}} - \frac{\partial_{a(b)}\alpha_k^{\text{aniso}*}}{\alpha_k^{\text{aniso}*}}\right], \nonumber \\
   &= n\Lambda_k^n \partial_{a(b)}[\ln \Lambda_k], \nonumber
   \end{align}
 we obtain
 \begin{align}
  \mathcal{F}_{ab}^{\text{aniso}} &= \frac{\Lambda_k}{(1-\Lambda_k)^2}[\partial_{a}\ln \Lambda_k\partial_{b}\ln \Lambda_k] , \nonumber \\
  &= \frac{4\Lambda_k}{(1-\Lambda_k)^2} \left[\partial_{a}\ln\left|\frac{\beta_k^{\text{aniso}}}{\alpha_k^{\text{aniso}}}\right| \partial_{b}\ln\left|\frac{\beta_k^{\text{aniso}}}{\alpha_k^{\text{aniso}}}\right|\right], \nonumber \\
  &= \frac{4\Lambda_k}{(1-\Lambda_k)^2}\left[\partial_{a}\ln\left|\frac{\beta_{k}^{(m)} + \beta_{k}^{(\xi)} + \beta_{k}^{(h)}}{1 + \alpha_k^{(m)} + \alpha_k^{(\xi)} + \alpha_k^{(h)}}\right|\partial_{b}\ln\left|\frac{\beta_{k}^{(m)} + \beta_{k}^{(\xi)} + \beta_{k}^{(h)}}{1 + \alpha_k^{(m)} + \alpha_k^{(\xi)} + \alpha_k^{(h)}}\right|\right].
 \end{align}
 This result, together with (\ref{betak}), allows us to determine the effects of the anisotropic perturbation on the estimation of the cosmological parameters. In order to perform the plots, let us assume that the wavevector $\overrightarrow{k} = (k_1, k_2, k_3)$ has a general direction specified by the spherical coordinates $(k, \theta, \phi)$ as
 \begin{align*}
 \overrightarrow{k} = k\sin\theta\cos\phi \hat{x} + k\sin\theta\sin\phi \hat{y} + \cos\theta \hat{z},
 \end{align*}
 where $k^2 = k_1^2 + k_2^2 + k_3^2$. Notice that the effects of the anisotropy is expected to depend on the direction of the particle momentum. Thus, the influence of the anisotropy on the Fisher information can be quantified by the azimuthal angle $\theta$ and/or the polar angle $\phi$.
 
\begin{figure}[H]
\centering
\includegraphics[scale=0.48]{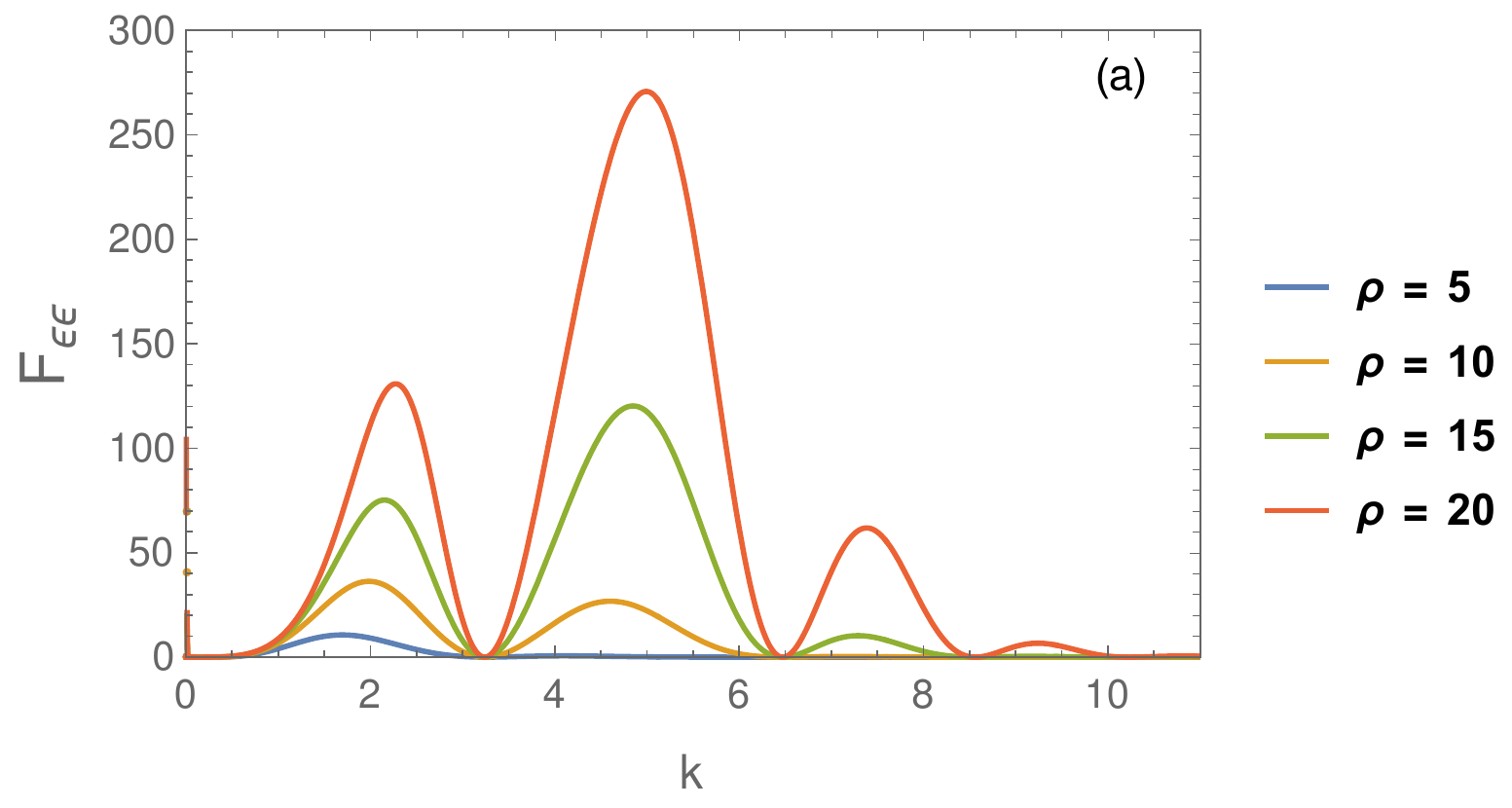} \includegraphics[scale=0.48]{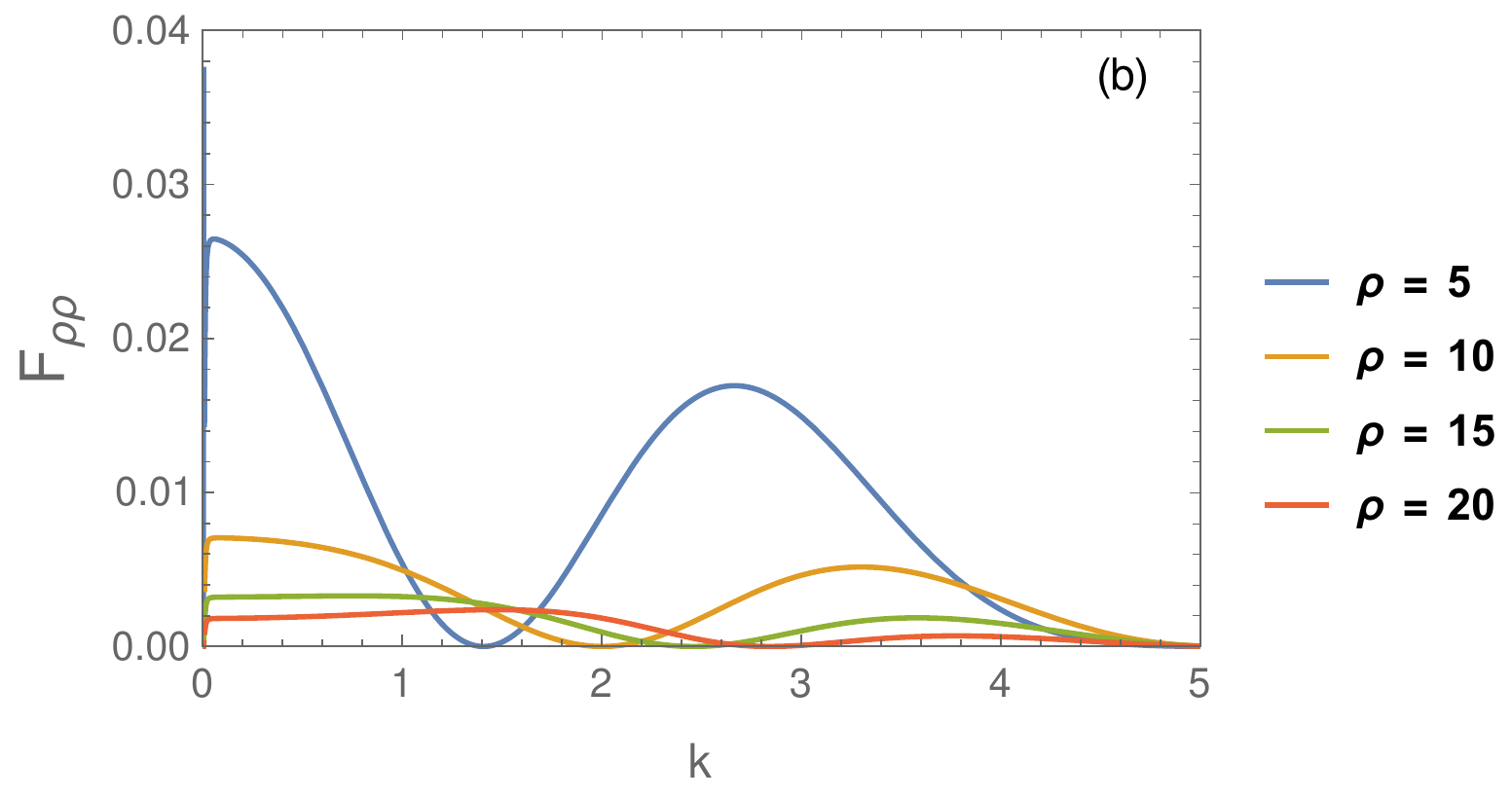} 
\caption{ Fisher information $\mathcal{F}_{\epsilon\epsilon}^{\text{aniso}}$ and $\mathcal{F}_{\rho\rho}^{\text{aniso}}$ as function of the momentum $k$ for different values of $\rho$. Here we have fixed $m = 0.001$, $\epsilon = 0.1$, $\xi = \frac{1}{6}$, $\theta = \phi = \frac{\pi}{2}$. }\label{FIaniso}
\end{figure}

  Note that the Fisher information $\mathcal{F}_{\epsilon\epsilon}^{aniso}$ and $\mathcal{F}_{\rho\rho}^{aniso}$ can be analytically evaluated, but it's not a convenient way. Of course, this can be analytically tackled under a particular approximation. However, let's do a numerical analysis of $\mathcal{F}_{\epsilon\epsilon}^{aniso}$ and $\mathcal{F}_{\rho\rho}^{aniso}$ without approximation. In Fig. \ref{FIaniso} we plot $\mathcal{F}_{\epsilon\epsilon}^{aniso}$ and $\mathcal{F}_{\rho\rho}^{aniso}$ as a function of the momentum of the mode $k$ for different values of expansion rate $\rho$. The plots show that $\mathcal{F}_{\epsilon\epsilon}^{aniso}$ and $\mathcal{F}_{\rho\rho}^{aniso}$ are oscillating functions of the momentum $k$. In particular, in Fig. \ref{FIaniso} (a) we observe that the oscillation of $\mathcal{F}_{\epsilon\epsilon}^{aniso}$ increases when $\rho$ grows. On the other hand, the Fig. \ref{FIaniso} (b) shows that the oscillation of $\mathcal{F}_{\rho\rho}^{aniso}$ decreases when the expansion rate grows. In addition, it is worth mentioning that, unlike of the isotropic case, the Fisher information spectrum for the estimation of the expansion rate $\rho$ is different of that for the estimation of the expansion volume $\epsilon$.

\begin{figure}[H]
\centering
\includegraphics[scale=0.5]{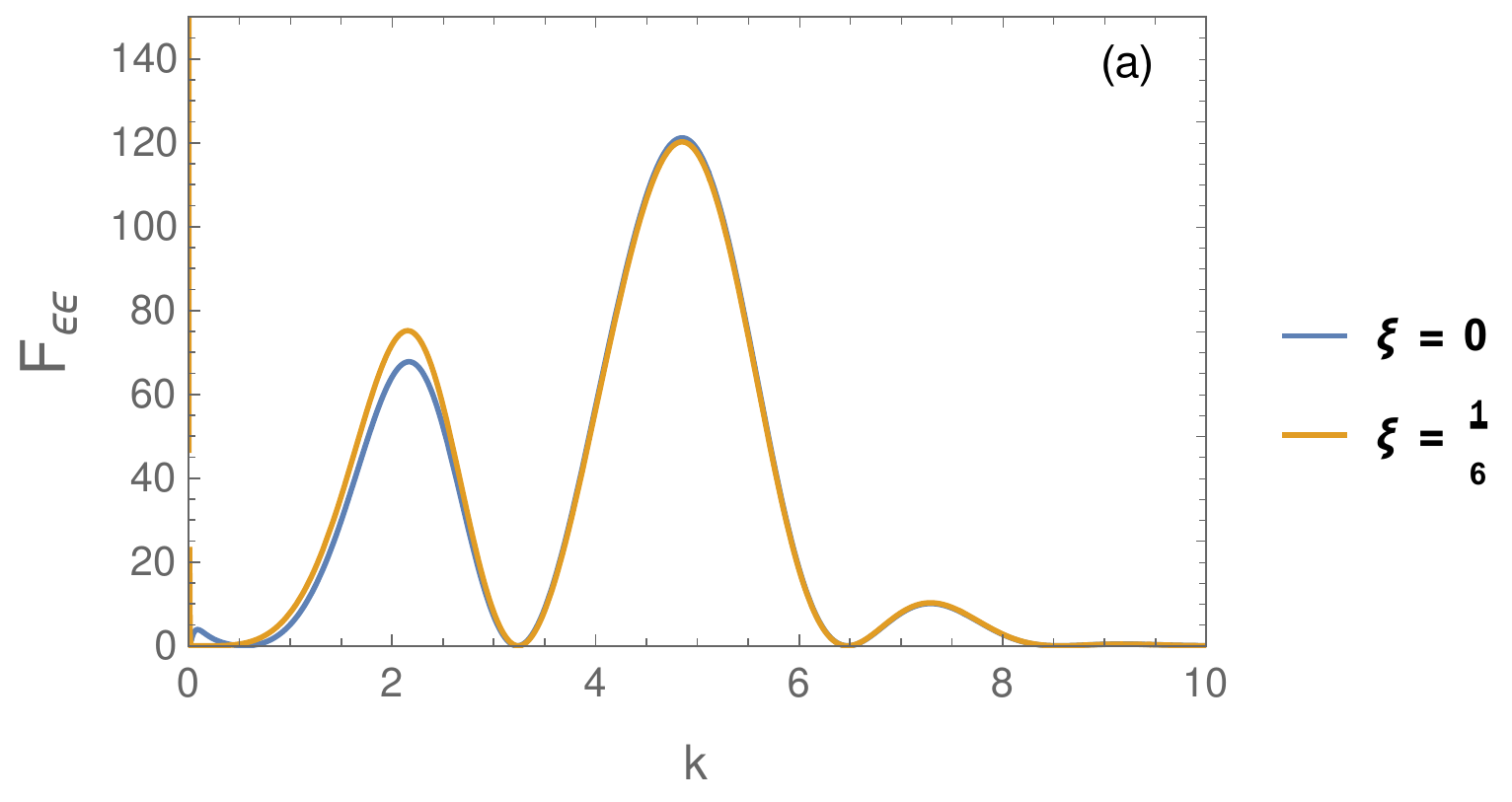} \includegraphics[scale=0.5]{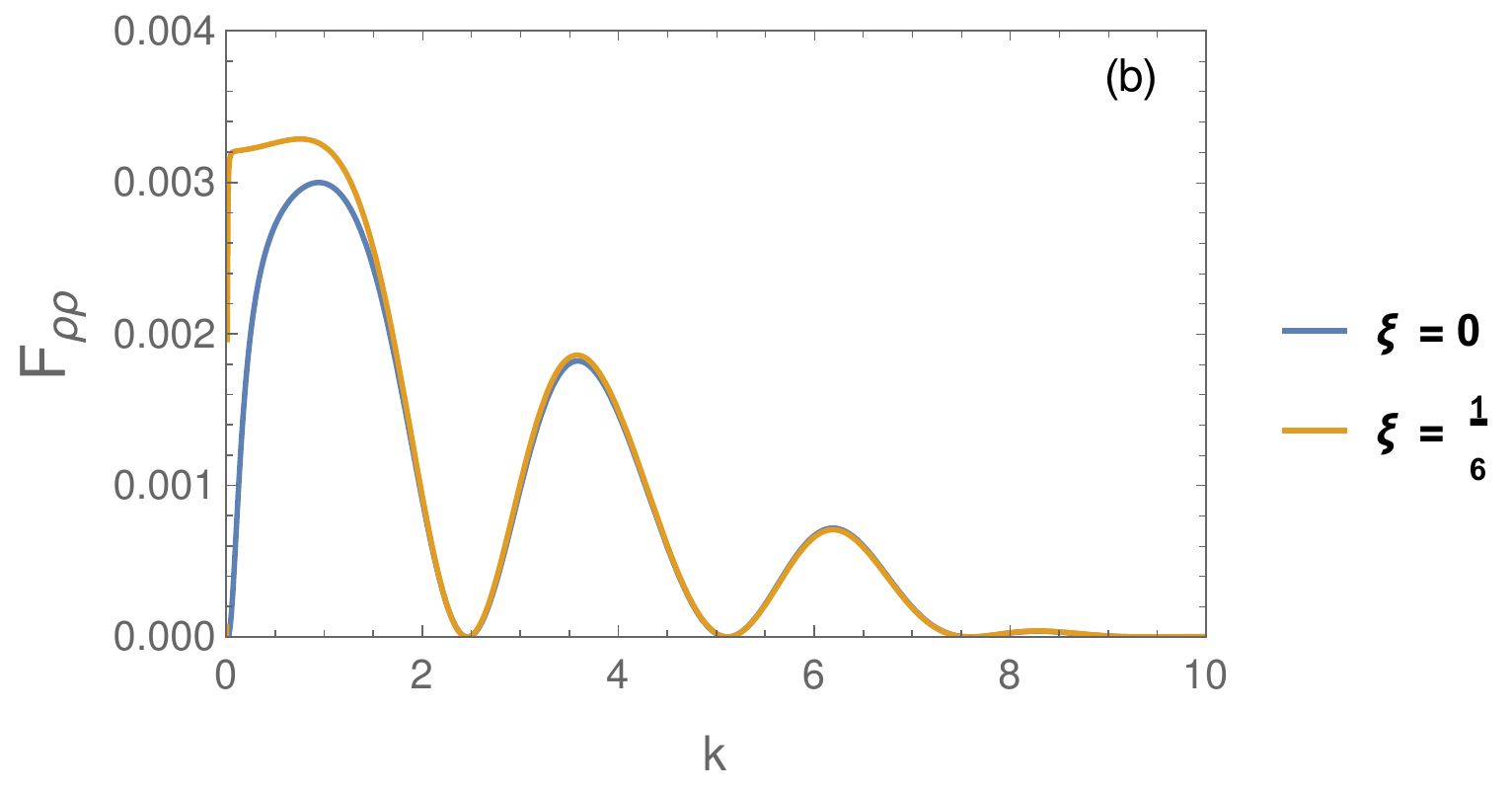} 
\caption{ Fisher information $\mathcal{F}_{\epsilon\epsilon}^{\text{aniso}}$ and $\mathcal{F}_{\rho\rho}^{\text{aniso}}$ as function of the momentum $k$ for different values of $\xi$. Here we have fixed $m = 0.001$, $\epsilon = 0.1$, $\rho = 15$, $\theta = \phi = \frac{\pi}{2}$. }\label{FIaniso2}
\end{figure}

 We are also interested in knowing how the coupling between the field and the spacetime curvature affects the estimation of the cosmological parameters. In Fig. \ref{FIaniso2} we plot $\mathcal{F}_{\epsilon\epsilon}^{aniso}$ and $\mathcal{F}_{\rho\rho}^{aniso}$ as a function of the momentum of the mode $k$ for different values of the paraneter $\xi$. Fig. \ref{FIaniso2} shows that, in low energy regime (for smaller values of momentum $k$), there is a difference in the behavior of the Fisher information $\mathcal{F}_{\epsilon\epsilon}^{aniso}$ and $\mathcal{F}_{\rho\rho}^{aniso}$ as the coupling $\xi$ changes from minimal ($\xi = 0$) to conformal ($\xi = 1/6$).

\begin{figure}[H]
\centering
\includegraphics[scale=0.5]{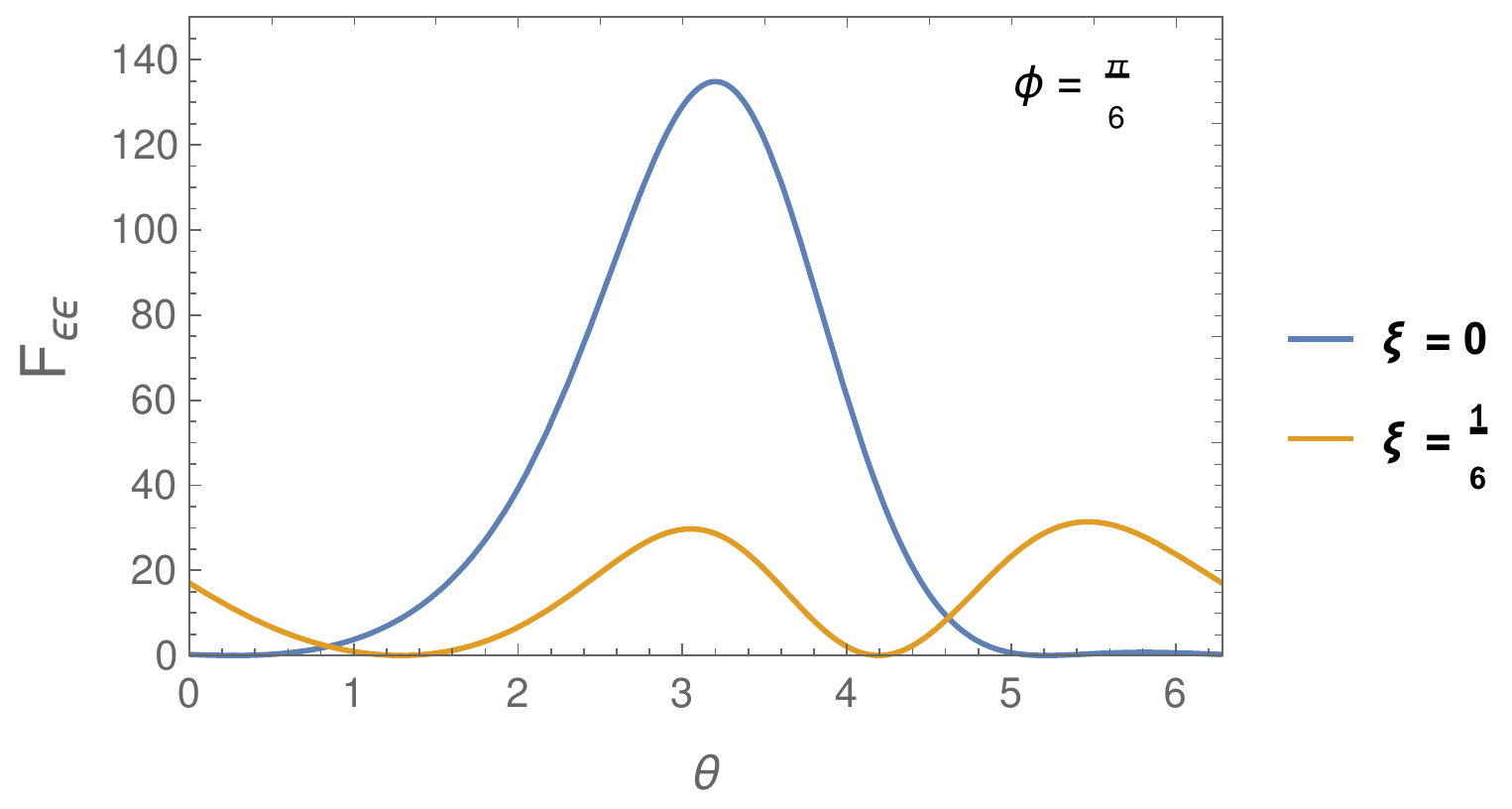} \includegraphics[scale=0.5]{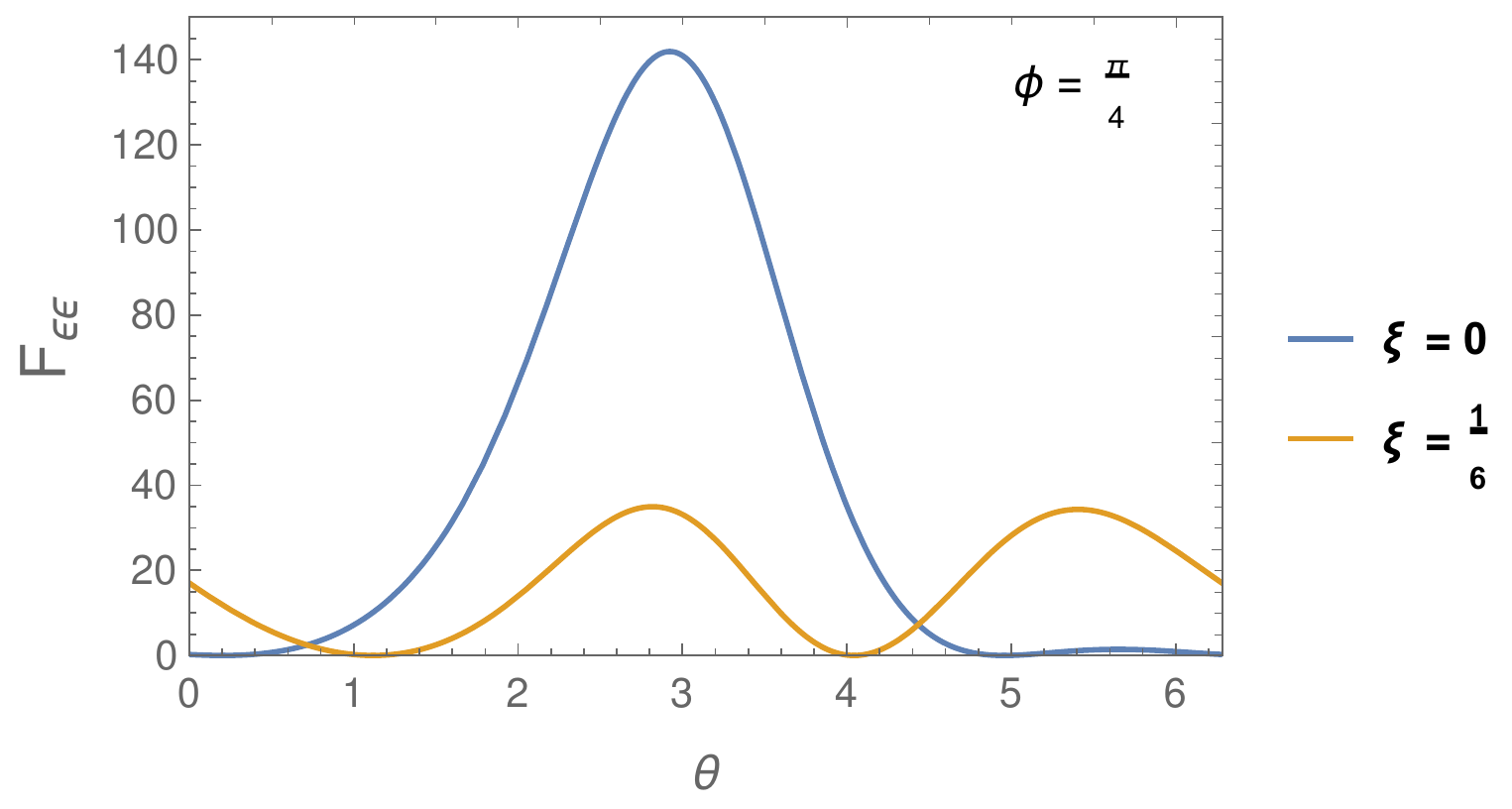} \includegraphics[scale=0.5]{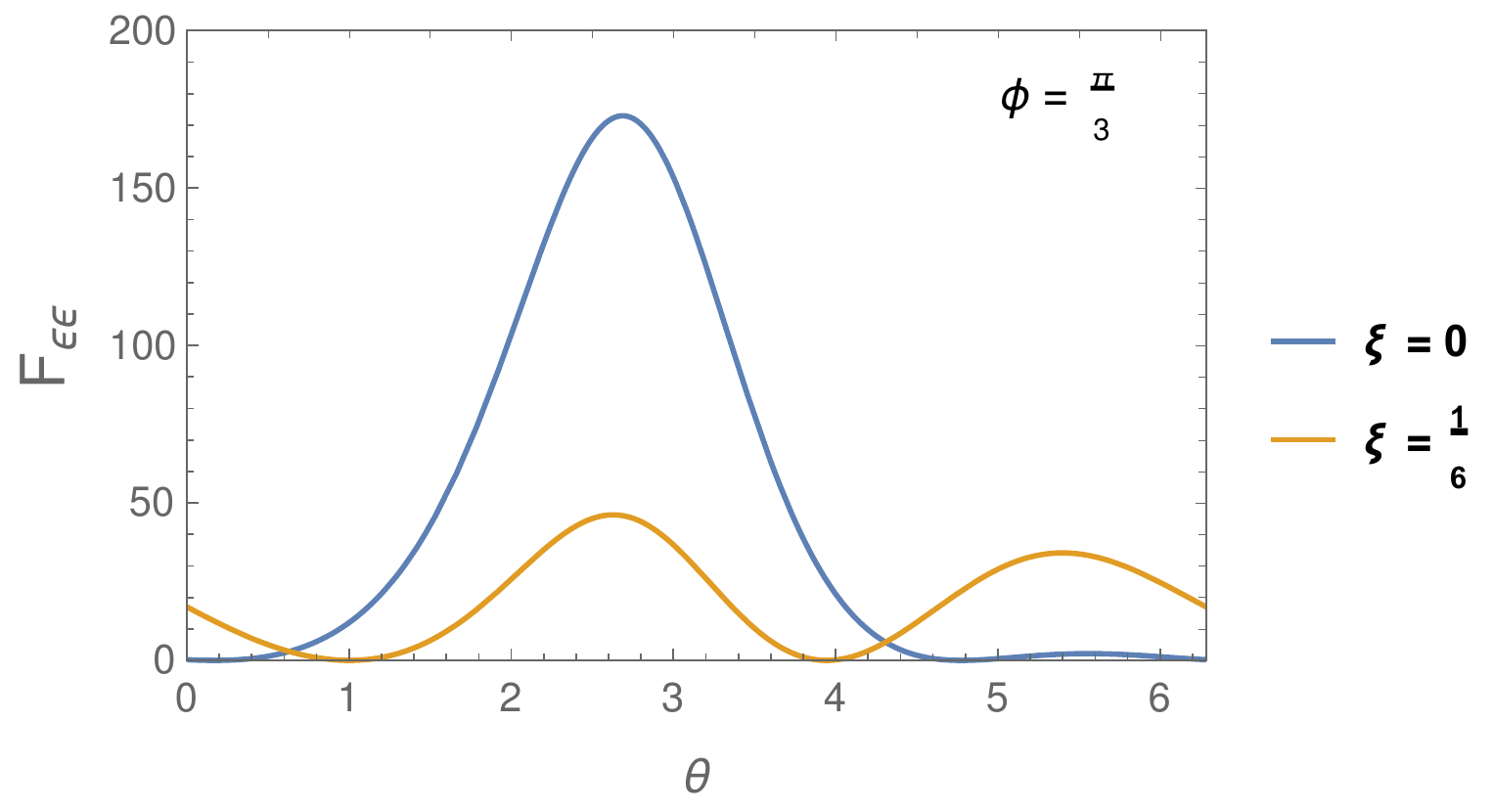} \includegraphics[scale=0.5]{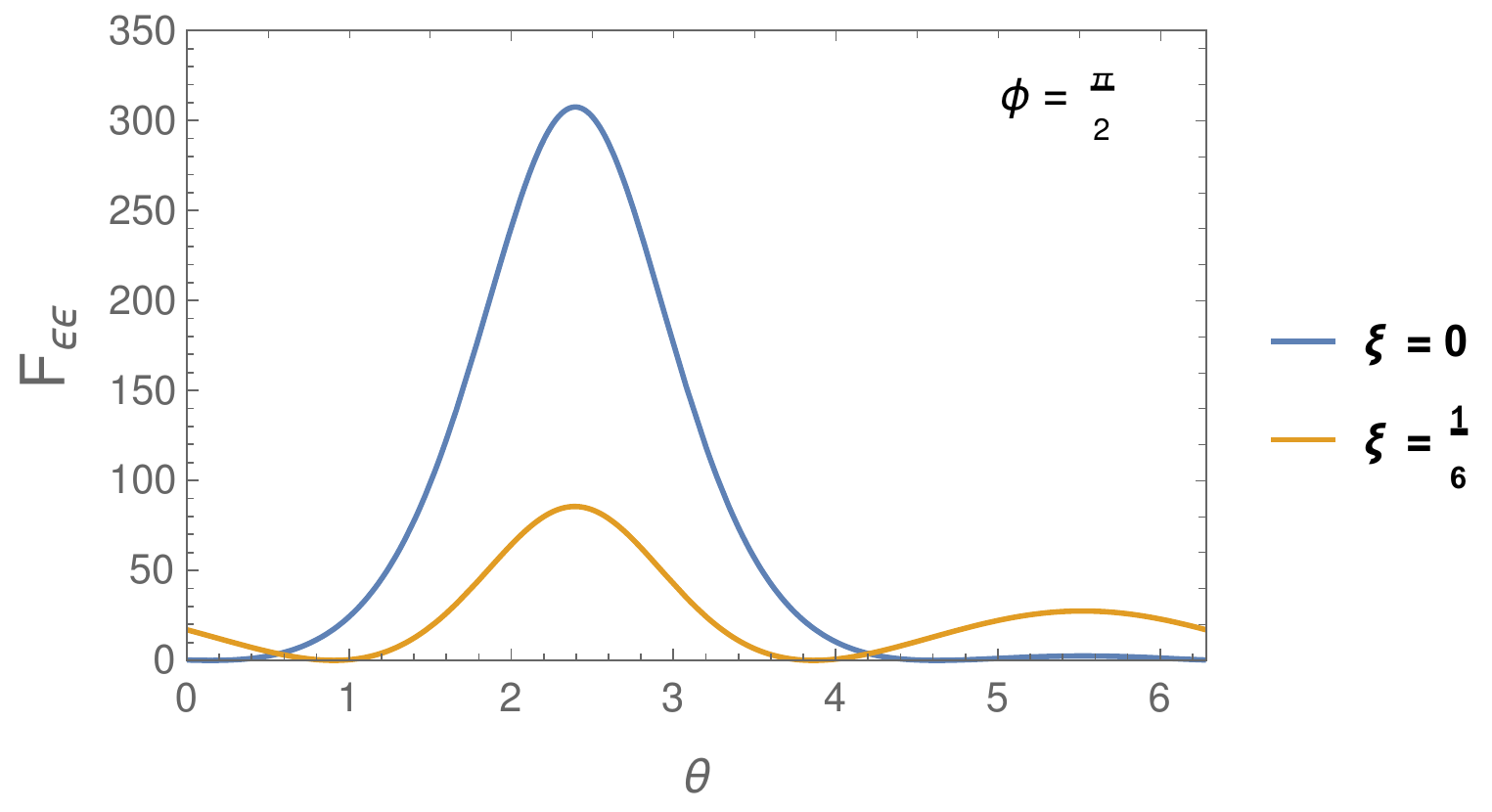}
\includegraphics[scale=0.5]{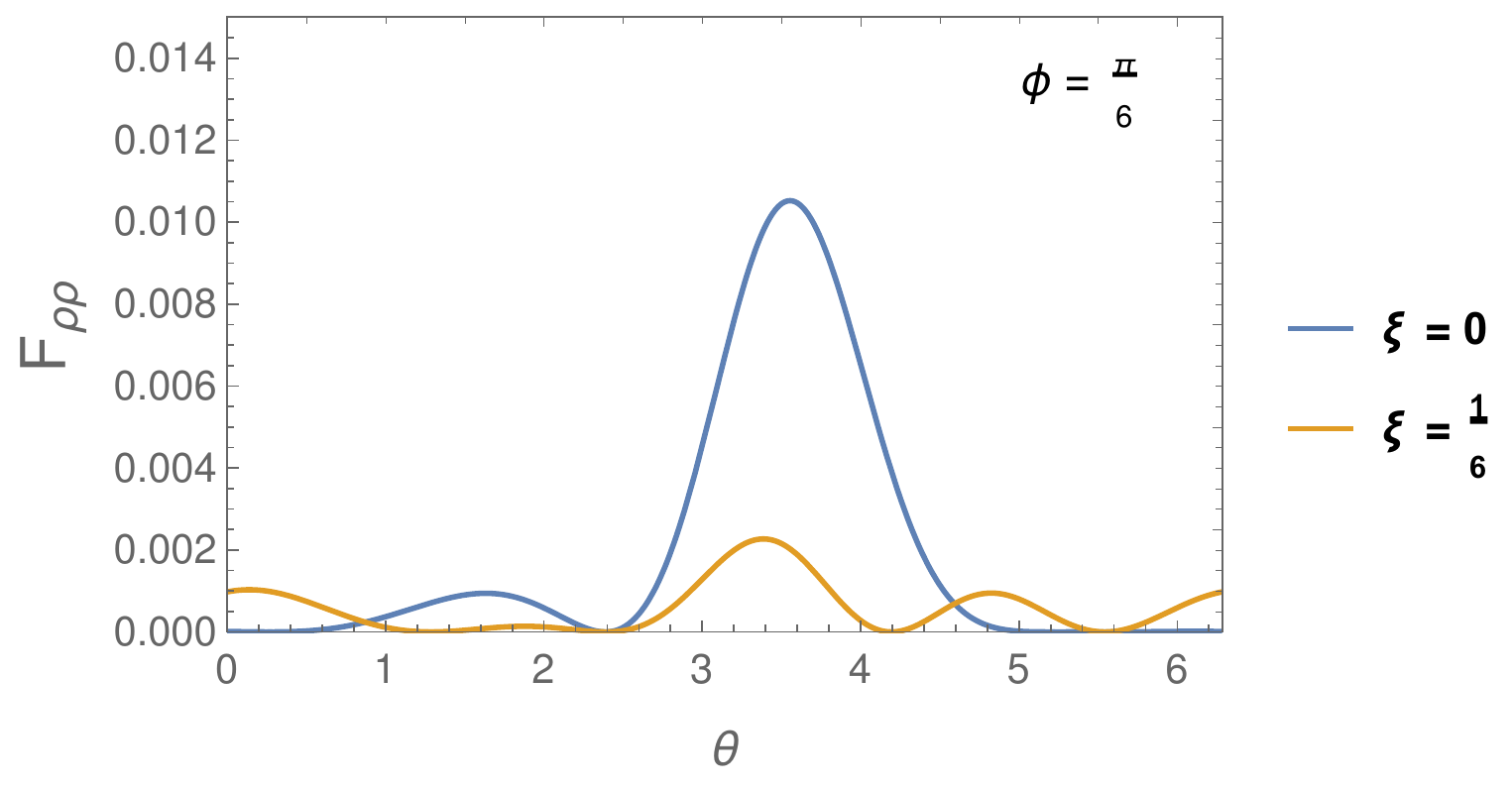} \includegraphics[scale=0.5]{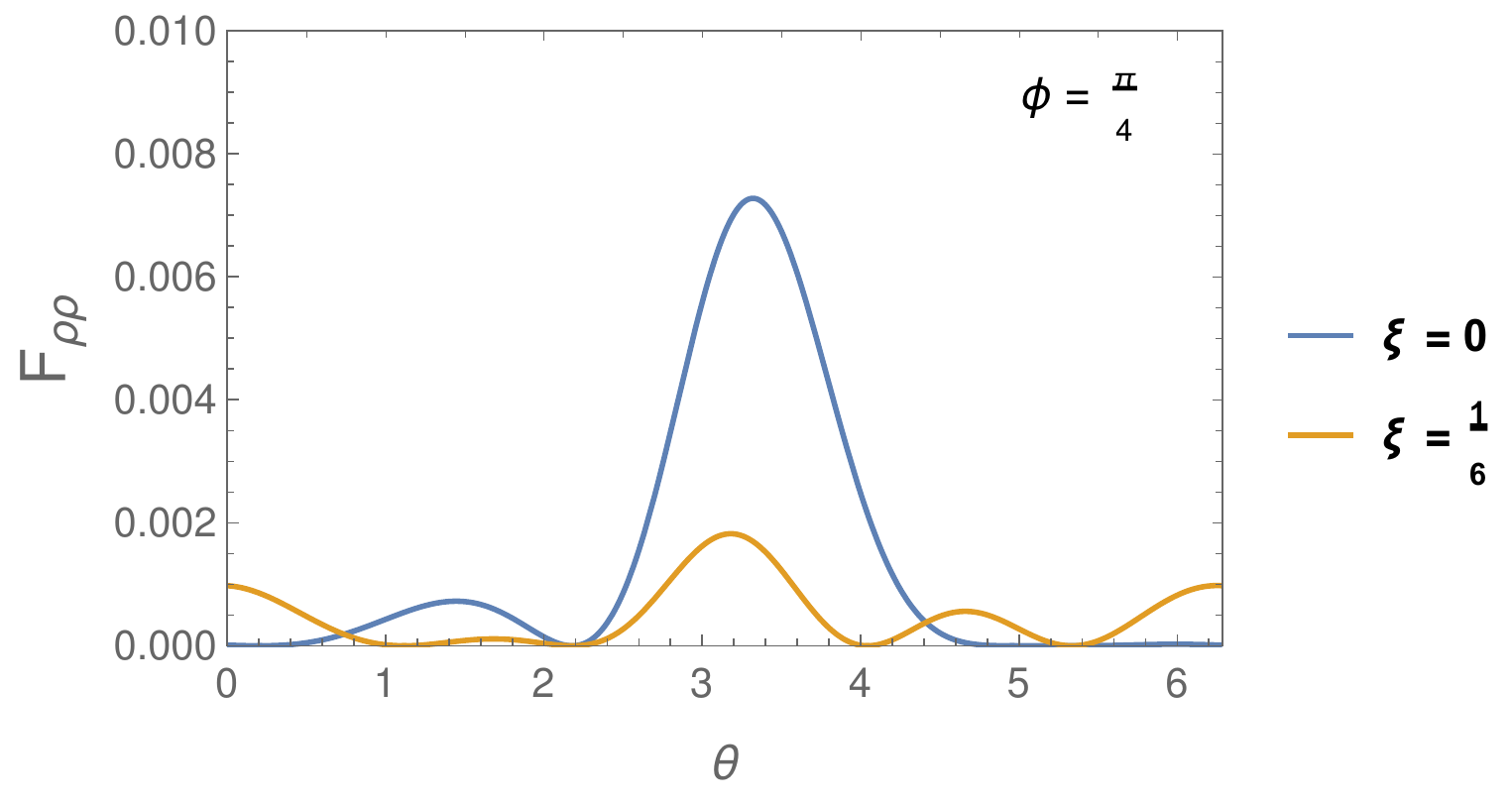} \includegraphics[scale=0.5]{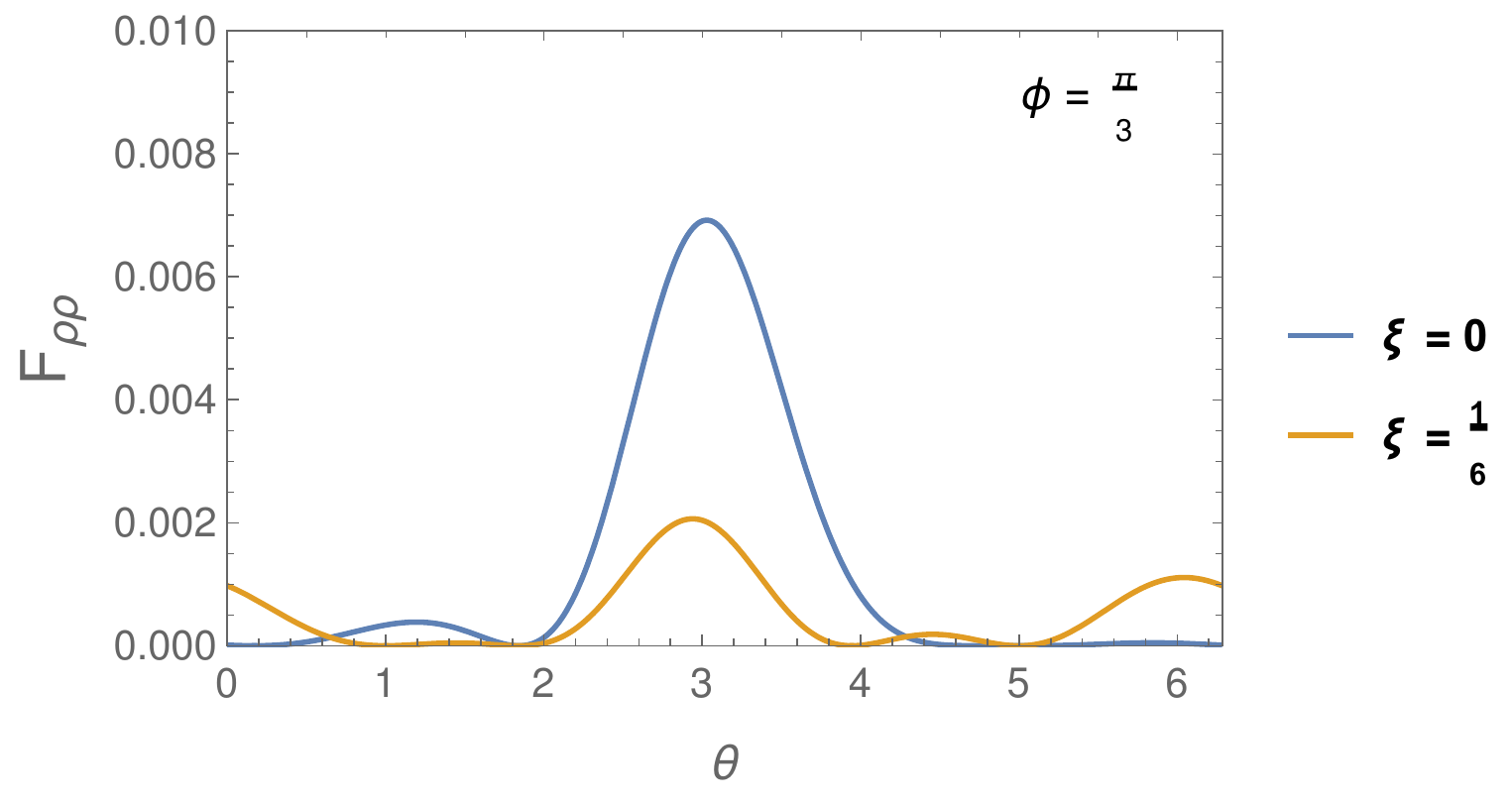} \includegraphics[scale=0.5]{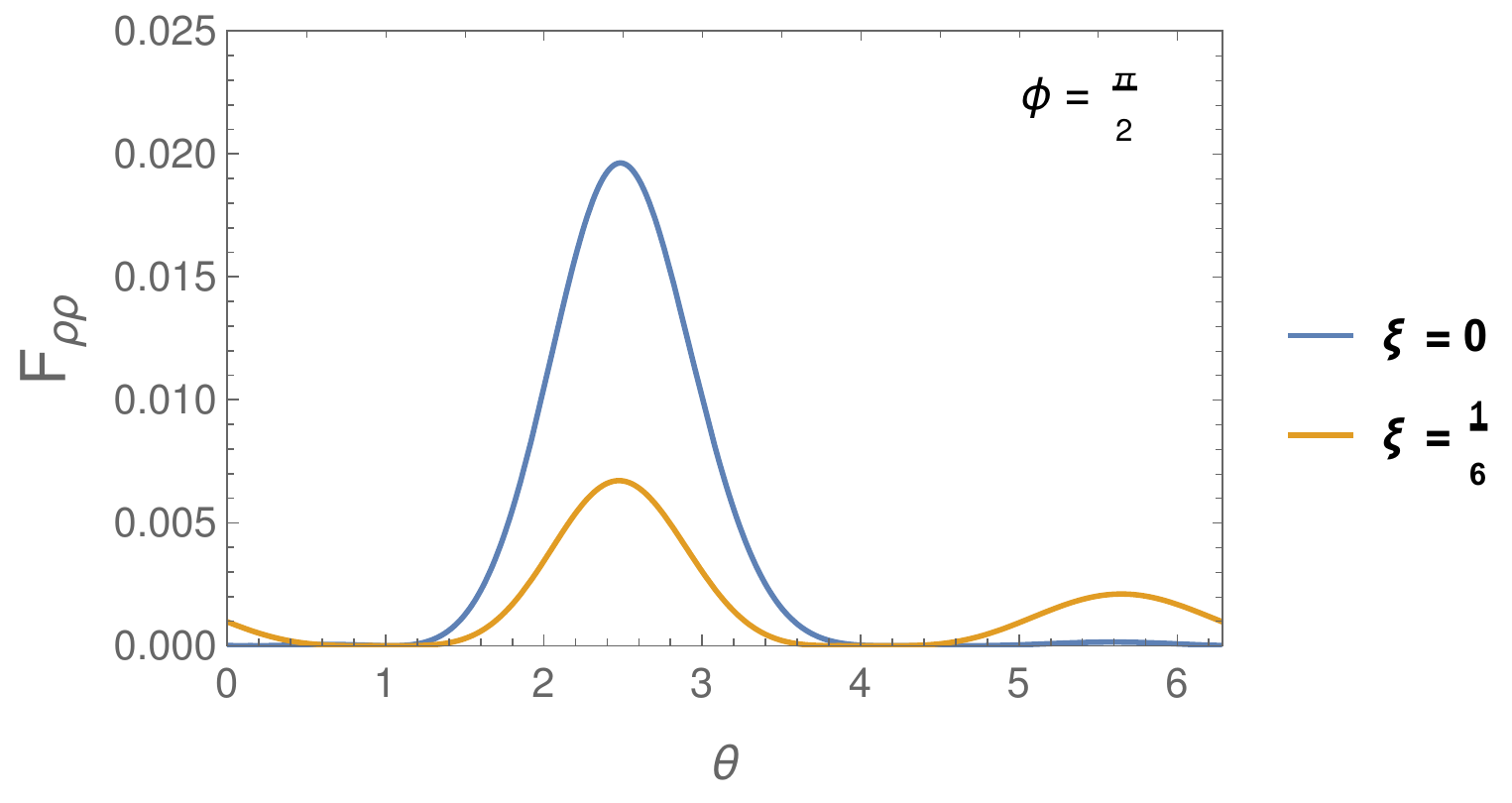}
\caption{ Fisher information $\mathcal{F}_{\epsilon\epsilon}^{\text{aniso}}$ and $\mathcal{F}_{\rho\rho}^{\text{aniso}}$ as function of azimuthal angle $\theta$ for different values of $\phi$. Here we have fixed $k = m = 0.01$, $\epsilon = 0.1$, $\rho = 1$. }\label{FIaniso3}
\end{figure}

In Fig. \ref{FIaniso3} we plot $\mathcal{F}_{\epsilon\epsilon}^{aniso}$ and $\mathcal{F}_{\rho\rho}^{aniso}$ as a function of azimuthal angle $\theta$ for different values of $\phi$. Fig. \ref{FIaniso3} shows that the estimation of the expansion rate $\rho$ and the expansion volume $\epsilon$ is sensible to the direction of the momentum $k$. For $\xi = 0$ the highest precision in the estimation of the expansion rate and the expansion volume can be achieved at a certain value of the azimuthal angle $\theta$. On the other hand, when $\xi = 1/6$ we observe that $\mathcal{F}_{\epsilon\epsilon}^{aniso}$ and $\mathcal{F}_{\rho\rho}^{aniso}$ are oscillating functions of the azimuthal angle $\theta$. It is interesting to note that the oscillations change as $\phi$ varies from $\pi/6$ to $\pi/2$.

 The above results suggest that a small gravitational disturbance (anisotropy) changes the behavior of the Fisher information, introducing oscillations, when compared with the isotropic case. The estimation of the expansion rate $\rho$ and the expansion volume $\epsilon$ is sensible to the direction of the momentum $k$. Other interesting feature to note is that there is a substantial difference between the Fisher information spectrum for the minimum and conformal couplings. This differerence is significant only for small values of momentum $k$.

\section{Conclusions}

Starting from expanding spacetime models, we presented the estimate of the cosmological parameters for the isotropic and anisotropic models. We have observed that, unlike of the isotropic case, a small gravitational disturbance (anisotropy) introduces oscillations in the Fisher information spectrum. The estimation of the expansion rate $\rho$ and the expansion volume $\epsilon$ is sensible to the direction of the momentum $k$ of particles created during the cosmic expansion. Moreover, we have seen that there is a substantial difference between the Fisher information spectrum for the minimum and conformal couplings. 
Our analysis suggests that both anisotropy and coupling between the field and the spacetime curvature significantly affect the estimation of the cosmological parameters.

\section*{Acknowledgments}
OPSN wishes to acknowledge support by the EDITAL FAPEPI/MCT/CNPq No. 007/2018: Programa de Infraestrutura para Jovens Pesquisadores/Programa Primeiro Projetos (PPP). I.G.P. acknowledges Grant No. 307942/2019-8 from CNPq. PRSC would like to thank CNPq for grants: Universal-431727/2018 and Produtividade 307982/2019-0. HASC would like to thank the Brazilian funding agency CAPES for financial support.

\end{document}